# AI is ready, but are the students? From answer-making to sensemaking.

Antoinette Stone, Ruoning Zhang, Brian Utter
*Physics Department, University of California, Merced, 5200 Lake Road, Merced, California, 95343*

Students in introductory physics often struggle with interpreting kinematics graphs, especially when connecting graphical features to physical meaning. We investigated AI tool use in kinematic graphing analysis as an intervention to support active sensemaking. This study examines GPT-5 performance on the *Test of Understanding Graphs in Kinematics* (TUG-K) and the way students interacted with GPT-5 during introductory physics activities involving graphical analysis. For the first part of the study, GPT-5 responses were evaluated across repeated TUG-K trials and compared to previously reported GPT-4 performance. Following that analysis, student-GPT-5 dialogue episodes were collected and analyzed using an inductive coding scheme. GPT-5 performance results show markedly improved accuracy. Student-GPT-5 dialogue episodes evolved from mostly procedural answer-making requests early in the term, to more reflective sense-making interactions at the end of the term, suggesting that instructional framing is needed for AI tools to function as meaningful peer-like learning supports.

# I. INTRODUCTION

Why are kinematics graphs so hard to analyze for students new to studying physics? In a recent kinematics graphing problem on a midterm exam at our institution, students performed poorly in connecting physical features of the graph to descriptions of the motion. Such performance in graphical analysis is not an isolated case, as many researchers have observed and reported on the numerous misconceptions and missteps that students encounter during their first exposure to the motion analysis activities introduced in an introductory calculus-based physics class. These obstacles and difficulties are intensified when analysis requires synthesizing graphical information, presenting significant pedagogical challenges associated with navigating these learning curves.

We looked to the current expansion of AI tool use in education to inform useful interventions to improve student outcomes in kinematics graphing analysis. We sought to leverage students' willingness to initiate AI as a resource for solutions steps, explanations, and immediate feedback on answer correctness [1]. This paper examines the use of GPT-5 tools in collaborative classroom environments, evaluating the efficacy of such tools for accuracy and peer-like learning support and assessing the level of student cognitive engagement based on linguistic cues.

# II. BACKGROUND AND RESEARCH GOALS

To assist in developing robust research questions we briefly summarize relevant background and previous work. In so doing, we address prominent ideas in two specific areas: 1) what prior research exists to explain the difficulties students encounter in graph analysis and 2) what previous work with AI tools provide foundational knowledge upon which to build an AI intervention. Many studies [2,3,4] that explore student difficulties with graphical analysis demonstrate that students frequently misinterpret kinematics graphs. Misconceptions include treating graphs as literal pictures of motion; conflating height and physical quantities; treating slopes as heights; misinterpreting areas under curves; and failing to recognize the distinction between instantaneous and average quantities. Some previous research examining AI tool use suggests such tools show promise for enhancing engagement, along with the positive effects of timely and targeted feedback [5]. Nevertheless, concerns remain, including overreliance and the negative effects of reducing the productive struggle needed for deep learning [6]. Moreover, the overreliance factor is made worse if the AI responses are in error, potentially reinforcing misconceptions that students are seeking to eliminate. This further suggests that the implementation of AI as peer-like learning supports tools, depends critically on accuracy as well as on instructional framing, determining whether they are positioned as answer-generating systems or as scaffolds for sense making and conceptual reasoning.

Recent research showed that ChatGPT-4 performs at a level comparable to an average high school student when presented with the "Test of Understanding of Graphs – Kinematics" (TUG-K) [7]. With the emergence of more advanced models such as GPT-5, it remains unclear whether advancing accuracy alone will provide active sensemaking peer-like learning support from the use of these tools. From this perspective we focus on uncovering cognitive cues from the student-GPT-5 interactions. To do this we needed to first establish a stable and accurate performance with GPT-5.

The purpose of this study is twofold; first to examine GPT-5's performance accuracy on the TUG-K as a prerequisite to effective student support, to then enable us to analyze student-AI interactions while using this tool across three specific physics units that made use of graphical analysis. Using a structured coding framework, we sought to find evidence to determine if GPT-5 was functioning as a peer-like learning support tool for active sensemaking. The study, then, is guided by these research questions: 1) How does GPT-5 perform on the TUG-K? 2) What cognitive activities are observed in student-GPT-5 interactions?

# III. METHODOLOGY

To address how GPT-5 performs on the TUG-K, we replicated a previous study which investigated GPT-4 performance on the TUG-K [8], utilizing GPT-5 for comparison. Since the general conclusion from this prior investigation determined that GPT-4 was not a good form of peer support, to proceed with the second research question, the utility of GPT-5 needed to be determined. To address the second research question, we evaluated student interactions with GPT-5. The students who participated were enrolled in large enrollment calculus-based, introductory physics class, in the spring term of 2026, which included lectures twice weekly along with a single weekly discussion session that spanned 1 hour-50 minutes. From this cohort, we collected data from several discussion sessions whose topics involved some form of graphical analysis. Our final data set came from a whole class analysis of student-GPT-5 dialogue episodes submitted as part of an end-of-term assignment. The data sets collected are described in section IV.

## A. ChatGPT performance study

To evaluate GPT-5 performance we presented each of the 26 TUG-K questions used in the prior study to the model 60 independent times. Each trial was conducted in a new conversation session with memory disabled to prevent information carryover between trials. The temperature was left at the model's default setting as in the first investigation. The same prompt was used for every test as follows: "Please read the question and give me the answer with 2 sentences of reasoning as to why you chose that answer".

Both the selected answer and the generated explanation were recorded for potential future work, but we focus on response accuracy for the present study. Differences between GPT-4.1 and GPT-5 performance on the TUG-K were analyzed using paired comparisons at the question level. Results of a paired sample t-test was conducted on item correctness for the same set of questions.

### B. Student-AI interaction study

To address the second research question, we investigated how students used AI support in their discussion sessions which were associated with the large enrollment, calculus-based, first-semester physics class in mechanics. The majority of the cohort were second-year bioscience students. The course structure included three discussion sections with equal enrollments of approximately 40 students with each section employing active engagement learning strategies where students collaborated in small groups. In each section they were tasked with working on strategies and approaches for solving problems that were provided in group worksheet format. Students were encouraged to access course materials, lecture notes, and other outside resources in their problem-solving activities.

AI use was allowed in all three discussion sections, however, we focus on a single section in which students were encouraged to use GPT-5 as a peer-like learning support tool during discussion activities. The section, with an enrollment of 43 students, was facilitated by 1 experienced Teaching Assistant (TA) and 4 Learning Assistants (LAs). In support of this activity, the TA provided general guidance for how to submit GPT-5 inquiries by implementing the following general guidelines: 1) Discourage inquiries that merely input question prompts or problem statements as written. 2) Assist and encourage students with reformulating questions. 3) Ask students to assess what they learned and to explain the AI response. In the remaining two discussion sections, students participated in traditional discussion sessions with AI use permitted but without any formal AI query guidance.

Three topics were selected for data collection: kinematics graphs and projectile motion, both topics occurring early in the term, and Hooke's Law, a topic occurring midway in the term. These were selected because each require graphical analysis, such as analyzing slopes, area under the curve, or other features in pursuit of solutions for their assigned discussion problem. At the end of each of the three sessions whose primary task was graph analysis, students submitted screenshots of their GPT-5 interactions, which we hereby refer to as dialogue episodes, to inform an inductive coding rubric.

An example of a discussion problem that students worked on is given in Fig. 1. At the end of the term, we again evaluated how students used AI support to complete an end-of-term assignment that involved a targeted analysis of a particular kinematic graph as shown in Fig. 2.

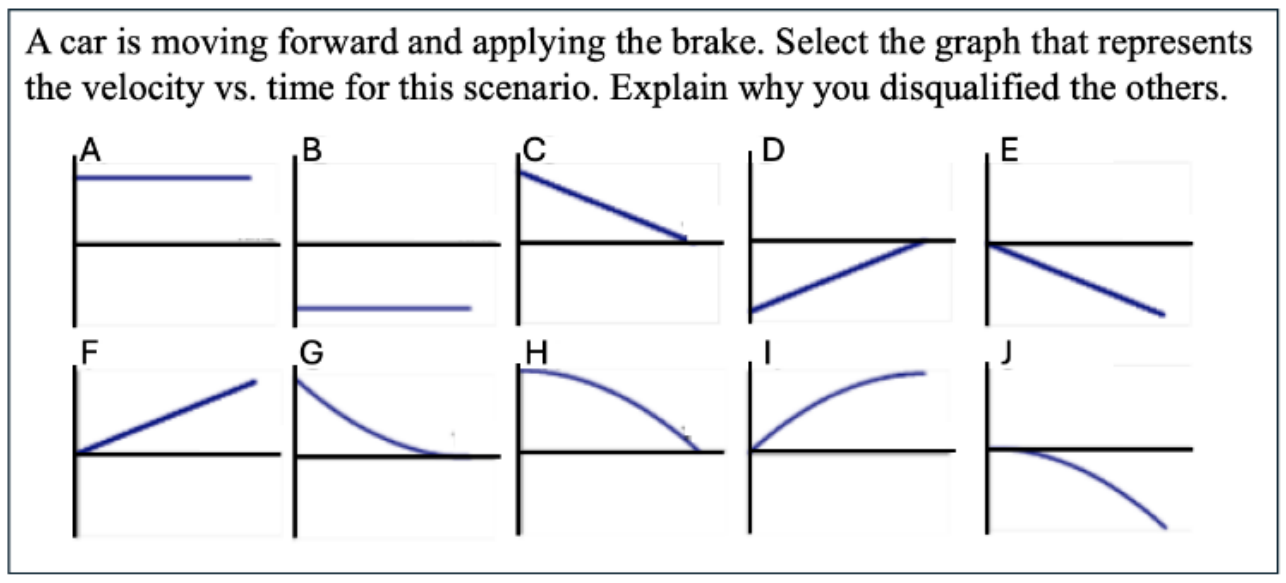

FIG. 1 An example of a kinematic discussion problem.

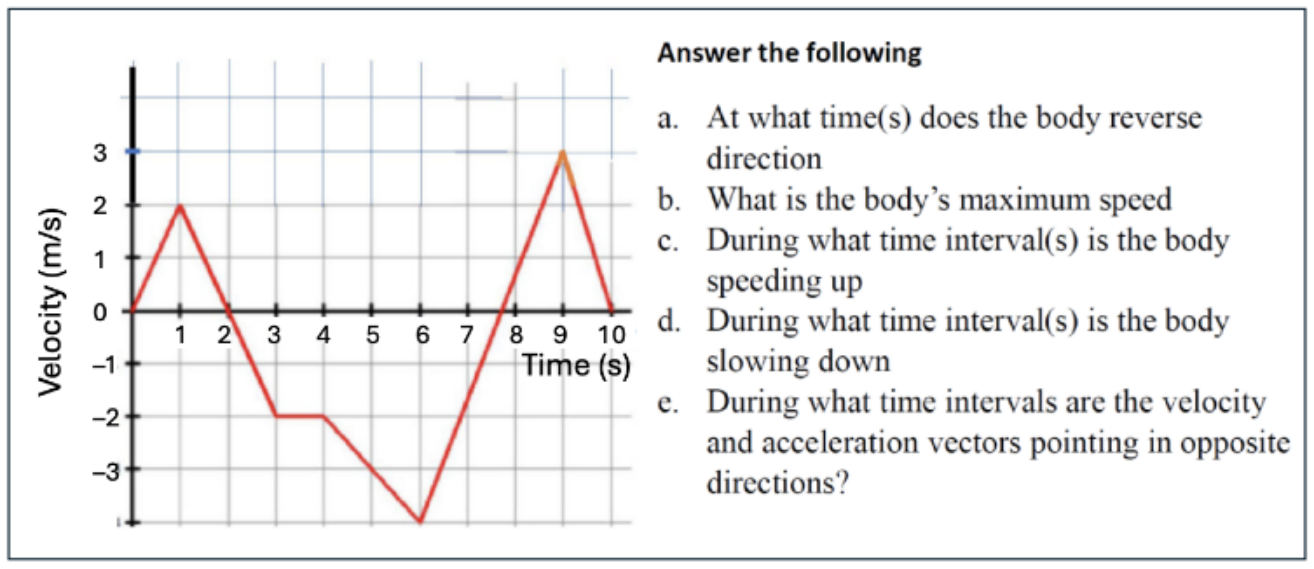

FIG. 2 End-of-term assignment.

## IV. DATA ANALYSIS

### A. GPT-5 Performance

Figure 3 shows GPT-5 outperformed GPT-4 on most TUG-K items. For instance, Item 24 (red) improved from correct on 1.6% of attempts in GPT-4 to 100% correct on GPT-5. This item is one of 15 that were answered 100% correctly in GPT-5 The mean improvement in item correctness was 43% (SD=39) to 84%. A paired-samples t-test comparing item correctness between models showed GPT-5 performance significantly higher than GPT-4, $t(25) = 5.64$, a corresponding p value $< .001$, and Cohen's d = 1.11, providing evidence that GPT-5 achieved significantly greater accuracy. Item 14 (blue) clearly deviated from this trend, decreasing from 81.6% to 40% correct in GPT-5. The incorrect response from GPT-5 as compared to GPT-4 for this item appears to be the result of GPT-5's inaccurate "reading" of the graphical selections presented as the solution set choices, consistent with the previous study's conclusion regarding the reasons for incorrect answers given by GPT-4 [8, 9]. We anticipate the rapid rate of improvement to continue, but find GPT-5 is sufficiently accurate to enable productive interventions on graphical problem solving.

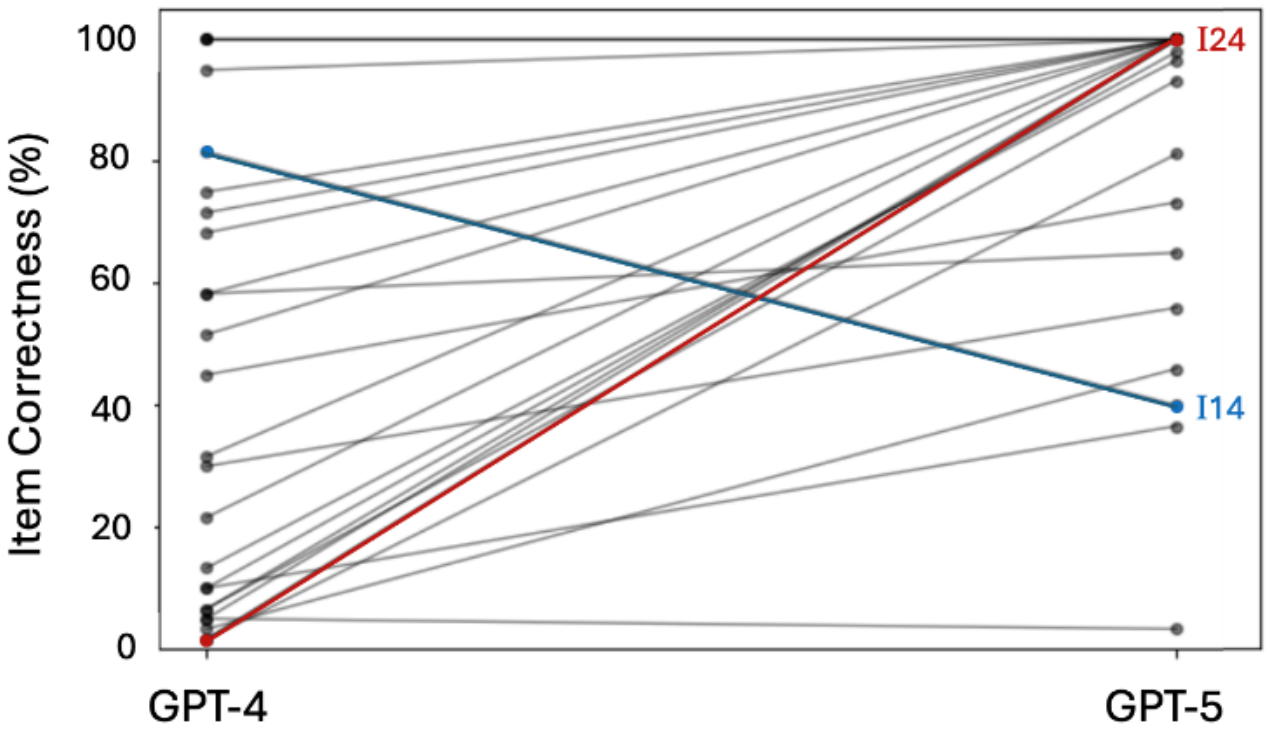

FIG 3. Item comparison of GPT-4 and GPT-5 performance on TUG-K. Each point represents one of the 26 items. Lines connect same item across models showing item specific changes. For example, Item 24 (red) correctness was 1.6% in GPT-4 and 100% in GPT-5.

### B. Developing the coding rubric

Coding student dialogue or texts for linguistic cues is a well-supported methodology for gaining insight into intent and meaning with a long history in science education. Since Lemke's seminal work, "learning science means learning to talk science" [10], researchers have long taken the position that students' language, gestures, and written texts mediate thought and intent into meaningful expressions and are evidence of cognitive activity [12-15] and it is this practice that motivated our analysis approach.

Screenshots of the student-AI dialogue transcripts were imported into Delve, a qualitative data analysis platform, where an inductive coding process was used to identify recurring patterns and themes. We identified repeated patterns of syntax and lexicon in pursuit of an inductive coding scheme resulting in the coding framework described in Table I, which shows categories assigned to the specific repeated texts and phrases that students used when accessing GPT-5 for peer-like learning assistance in their collaborative problem-solving efforts during discussion. The collected samples included 134 student-GPT dialogue episodes via screenshots over these three different graphing topics.

We now show examples of the explicit texts and phrases submitted which provided the linguistic cues to support our characterizations of the coding categories. For example, phrases such as

"*Show me step-by-step for how to go about this*" or "*give me a procedure without giving me the answer*" or "*walk me through this multiple step question and guide me through it*";

were repeated often, indicating efforts to seek recipes. Although these requests avoided simple direct answer requests, they clearly differed from those texts that sought to prioritize, compare, or differentiate relationships. For example, inquiry phrases such as

"*How is the spring constant related to the slope*" or "*Why should the area method give the same answer as the equation method?*"

that explicitly use the word "relate" and "same" to provide reasonable linguistic evidence of efforts to relate, compare or differentiate physical features, variables or relationships. We consider these inquiries as opportunities students take to "dig deeper" for understanding, or to reframe a problem in a way that helps them "make sense" therefore treating them as reasonable evidence of sense-making rather than answer-seeking actions [16]. Other texts demonstrated clear attempts to make logical connections, through cause-and-effect observations, as noted by the use of "if-then," or "when-then" statements. For example, statements such as

"*If the net force is zero then how are the applied force and frictional force related?*" or "*if the area is below the horizontal would that still be the work done?*", clear if-then examples, were considered as epistemological attempts for framing the problem which point to robust efforts in sensemaking in contrast to answer making.

Table II shows the frequency of these patterns and repeated phrases collected from the end-of term assignment as compared to those submissions collected from the discussion session. Note the final additional category, not observed in the earlier interactions, discussed below.

TABLE I. Categories of text patterns and phrases used in discussion session dialogue episodes with GPT-5

| Category | Examples of repeated patterns or phrases |
|---|---|
| Seeking recipes | " Show me how to find.." "Give me the steps", Walk me through…." |
| Seeking confirmation | "Is this the right next step", "Is this answer correct?" |
| Identifying, prioritizing | "What is the most important thing to consider", "What should I look for first?" |
| Relating, comparing, differentiating | "How does X relate to Y?", "How is X different from Y?" |
| Causal reasoning through logical connections | "If X is true, then is Y true/not true?", "When X is true, does Y result?" |

TABLE II. Text pattern frequency for discussions and end of term assignment

| Category | % Occurrence in discussion session | % Occurrence in end-of-term assignment |
|---|---|---|
| Seeking Recipes | 52% | 22% |
| Seeking Confirmation | 26% | 16% |
| Identifying, prioritizing | 11% | 10% |
| Relating, comparing, differentiating | 7% | 15% |
| Causal reasoning through logical connections | 4% | 12% |
| Think aloud protocols and/or asking for explanations physical meaning | 0% | 25% |

While the end-of-term assignment was administered to the entire class, only those student submissions who were also in discussion section which implemented GPT-5 inquiry guidance were analyzed for comparison. Coding consistency was evaluated using percent agreement between two raters who independently coded the submissions. Each rater compared submitted texts for repeated patterns and phrases and categorizing them based on the coding framework. On initial follow-up the raters agreed on 95% of the coded segments and attained 100% agreement after simultaneously revisiting texts whose category assignments were not in agreement.

## V. DISCUSSION

We consider the categories of *Seeking Recipes or Confirmation* answer-making pursuits, while the remainder, *Identifying, prioritizing, relating,* etc., are more aligned with sense-making behaviors. The last category, *Causal Reasoning* is the strongest example of "making sense" through epistemological activity involving the construction of explanations and descriptions [16].

The analysis of the dialogue episodes collected from the discussion sessions showed that most interactions were requests for processes, procedures, or recipes, for the purpose of attaining roadmaps to answers. The coded texts similarly suggest that the dialogue episodes for all three of the graphical topics explored in the analyzed discussion session demonstrated that sense-making pursuits were superseded quite significantly by answer-making endeavors.

On the other hand, analysis of the dialogue episodes collected from the final project provided linguistic evidence that students were interacting with GPT in a more structured sensemaking fashion, using more causal reasoning and asking for explanations involving physical meanings. The most notable difference in the texts were multiple appearances of dialogue episodes that looked similar to "think-aloud" protocols; dialogue episodes whereby students would attempt to summarize, frame or reframe the problem and within the same text, make an assumption or suggest an idea, and then ask for follow on confirmation [17]. Consider, for example, the following pattern, seen often:

"*In this graph I'm able to see the velocity and time, however I'm not able to determine if it changes directions. I'll assume that the object is initially moving forward. Is it always like this?*

*So, are you telling me if the velocity proves to be negative the object is just going back from its original point? Oh, I see, so negative doesn't mean slowing down, but rather moving in the opposite direction, so every time the line crosses the horizontal axis, the object is moving in the reverse direction?"*

These observations uncover a pattern like Initiation-Response-Feedback (IRF) classroom dialogues, where the instructor initiates a prompt, the student responds, and the instructor responds with additional feedback [18,19]. This pattern only works when the initiation is an idea or assumption. So, if the student had merely requested an answer in a closed-ended format, such as "tell me how to find the slope" the dialogue would not have generated an IRF pattern. But as students started to initiate their dialogue episodes with their own ideas and concepts, a natural IRF pattern began to emerge.

## VI. CONCLUSIONS AND FUTURE WORK

We find GPT-5 capable of accurately interpreting and guiding students on graphical problems, a marked improvement over prior results reported in which GPT-4 was assessed at high school level proficiency and unable to effectively support student development. While this may be expected given the rapid development of generative AI, it provides an important snapshot of the current state of AI and grounding for the subsequent work on student learning support.

The dialogue episodes coded from the end-of-term project are encouraging. The linguistic cues support notably improved efforts to interact with GPT in a way that supports sensemaking. The dialogue also shows that the tools were used as effective peer-like learning supports when accessed in this manner. The remarkable case-in-points are the multiple appearances of IRF dialogue similar to talk aloud/think aloud protocols. This is significant, we think, for helping to design approaches for ways in which we frame or instruct our students in their use of AI tools [20].

Why did this happen at the end-of-term project and not in the discussion sessions earlier in the term? Clearly, the discussion sessions provided students with practice developing their dialogues. We note also that the end-of-term assignment contributed to their grade, creating a higher stakes environment for showing evidence of more effective engagement as compared to the discussion environment. They were also given a rubric with a clear expectation of how points were to be garnered and applied to their overall grade and were cautioned against merely copying the prompt into the GPT, instead guided by examples in the rubric encouraging them to use GPT in a dialogic way. Additionally, they were asked to show at least three iterative dialogue strings, most likely contributing to the more cohesive flow of discourse patterns that we observed.

Future comparison of the different levels of scaffolding support, improvements in learning gains, and the factors contributing to AI as a sensemaking tool identified above could help us understand how best to frame and support effective AI-student interactions. The current study however suggests that AI has reached a level of accuracy sufficient to support graphical analysis and that guidance for how to frame and use GPT can play a significant role in helping students develop the skill for interacting with GPT to support active sensemaking and promote utilization of this resource as a peer-like learning support tool.